\documentclass[times,twocolumn]{aastex63}
\usepackage{multirow,graphicx}

\hypersetup{linkcolor=red,citecolor=blue,filecolor=cyan,urlcolor=magenta}

\received{...}
\revised{...}
\accepted{....}
\submitjournal{ApJ}

\shorttitle{X-ray study of WR 121\rm{a}}
\shortauthors{Arora \& Pandey}

\begin{document}

\title{Unraveling the nature of a deeply embedded Wolf-Rayet star WR 121a}

\email{bharti@aries.res.in, jeewan@aries.res.in}

\author{Bharti Arora}
\affiliation{Aryabhatta Research Institute of Observational Sciences (ARIES), Nainital$-$263 002, India}
\affiliation{School of Studies in Physics \& Astrophysics, Pt. Ravishankar Shukla University, Raipur$-$492 010, India}

\author{J. C. Pandey}
\affiliation{Aryabhatta Research Institute of Observational Sciences (ARIES), Nainital$-$263 002, India}

\begin{abstract}

An X-ray study of a deeply embedded Wolf-Rayet star WR 121a has been carried out using long-term (spanning over $\sim$12 years) archival observations from \textit{Chandra} and \textit{XMM-Newton}.  For the first time, a periodic variation with a period of 4.1 days has been detected in the X-ray light curve of WR 121a. No companion is seen in a merged and exposure corrected \textit{Chandra} X-ray image of WR 121a as shown in the other previous observations in \textit{J}-band. The X-ray spectrum of WR 121a has been well-explained by a thermal plasma emission model with temperatures of 0.98 $\pm$ 0.34  keV and 3.55 $\pm$ 0.69 keV for the cool and hot components, respectively and non-solar abundances. The present study indicates that WR121a is one of the X-ray bright massive binaries with an X-ray luminosity of $\sim$10$^{34}$ erg s$^{-1}$, which can be explained by an active wind collision between its components. The phase-locked modulations have been seen in the flux variation of WR 121a where the flux is increased by a factor of $\sim$1.6 from minimum to maximum in 0.3-10.0 keV energy band. These variations could be caused by an eclipse of the wind collision region by the secondary star in a  binary orbit. The winds of both components of WR 121a appear to be radiative.  Radiative inhibition as well as radiative braking are the most likely processes those are affecting the wind collision  severely in this short period massive binary system.  

\end{abstract}
\keywords{Wolf-Rayet stars (1806), Early-type stars (430), Binary stars (154), Stellar winds (1636), X-ray stars (1823), WN stars (1805)}


\section{Introduction}
Hot stars with spectral type from early-B to O lie in the upper left part of the Hertzsprung$-$Russell diagram. The evolution of these objects happens over short time scales (typically of the order of a few to 10 Myr) and they appear as Wolf$-$Rayet (WR) stars during the end stages of their evolution. The typical initial mass of the progenitors of WR stars is greater than 25 M$_{\odot}$ \citep{2007ARA&A..45..177C}. WR stars are set apart by powerful outflows of matter from their surfaces fed by their extreme radiation fields in the form of stellar winds \citep{2008A&ARv..16..209P}. These stellar winds travel with large terminal velocities (1000$-$3000 km s$^{-1}$) and lead to enormous mass-loss rates (10$^{-7}$ to 10$^{-4}$ M$_{\odot}$ year$^{-1}$). The presence of these potent outflows gives rise to many observational signatures which can be observed over a wide range of the electromagnetic spectrum. 

Massive stars in our Galaxy are born pre-dominantly within the dense cores of giant molecular clouds. They start affecting their environment very soon after a star has formed through their intense ionizing radiation fields as well as strong stellar winds. W43 (=G30.8$-$0.2) is one such star forming complex, which was first discovered in radio domain by \citet{1958BAN....14..215W}. It lies in the first galactic quadrant with galactic longitude of 30.8$^\circ$ and latitude of  $-0.2^\circ$ along the connecting point of the Scutum-Centaurus arm and the Galactic bar. The distance estimated for the W43 complex is $\sim$ 6 kpc from the Sun and it contains a total mass of $\sim 7.1 \times 10 ^{6} M_{\odot}$ \citep{2011A&A...529A..41N}. The core of W43 hosts a well-known giant HII region powered by an open cluster of young and luminous stars embedded in it (named in the SIMBAD\footnote{https://simbad.u-strasbg.fr/simbad/  \label{footnote 1}} database as [BDC99] W43 cluster). The detailed study of this region was performed in infra-red wavelength by \citet{1999AJ....117.1392B} and they pointed out that W43 cluster is totally  obscured in optical band with an estimated value of $A_{V}$ of $\sim$ 34 mag, which is equivalent to a hydrogen column density of about 6.5 $\times$ 10$^{22}$ cm$^{-2}$ \citep{2011A&A...532A..92L}.  \citet{1999AJ....117.1392B} also showed that the three brightest members of this cluster are massive stars, one of those is a Wolf-Rayet (WR) star (named as W43 \#1 = WR 121a) and other two belong to the category of O-type giants$/$supergiants (named as W43 \#2 and \#3).  Using these spectral classifications, the dust extinction ($A_{K}$) to the individual object could be estimated as close to 3.5 mag for all of them.

WR 121a was classified as WN7+abs owing to the similarity of its $K-$band spectrum with that of WR 131 \citep{1999AJ....117.1392B}. The ``abs'' component was added to the spectral type of this object because of the presence of the diluted emission lines in its spectrum which might be originating from the unseen companion of this star.   In the 7$^{\rm th}$ catalogue of Galactic Wolf-Rayet stars, WR 121a  was placed in the category of probable binary systems and was specified with WN7+a$/$OB? spectral type  \citep{2001NewAR..45..135V}. WR 121a was seen in X-rays for the first time by \citet{2001ApJS..134...77S} during the \textit{ASCA} Galactic Plane Survey (AGPS). They detected this source (AX J184738$-$0156) in 0.7$-$10.0 keV energy band with  $>$5$\sigma$ significance level and it was unidentified at that time. \citet{2007A&A...462L..37G} also pointed hints towards the binary nature of WR 121a using the spectro-astrometric technique. \citet{2011ApJ...727..105A} investigated the previously unidentified X-ray sources during AGPS with \textit{Chandra X-ray Observatory} data. They pointed out that the source AX J184738$-$0156 is co-incident with WR 121a discovered by \citet{1999AJ....117.1392B}. Based upon the \textit{Chandra} and \textit{XMM$-$Newton} spectral features of WR 121a,  \citet{2011ApJ...727..105A} suggested that this object belongs to a class of colliding wind binaries. However, they did not find any hint of variability in the data. Later, in the same year, \citet{2011A&A...532A..92L} studied the infra-red and radio properties of the W43 cluster. They reported that  two components (W43 \#1a and W43 \#1b) of WR 121a has been resolved with a separation of 598 $\pm $ 3 mas and a position angle of 255$^{\circ} \pm$ 1$^{\circ}$ using \textit{J}-band observations from ESO Very Large Telescope (VLT). Based upon the position of  W43 \#1b which was found to be exactly coincident with the position of an X-ray source CXO J184736.6-015633, they have also speculated that W43 \#1a is likely the WR member of the system, while W43 \#1b is the O-type companion since the X-ray emitting region generally lies close to the star with relatively weaker wind in a colliding wind binary.  Non-thermal synchrotron radiation was also  found to emerge from an extended region of the cluster but with a peak offset from the position of WR 121a by $\sim$3\arcsec (much closer to W43 \#3). As suggested by \citet{2011A&A...532A..92L}, this non-thermal radio emission is originating due to the cumulative effect of stellar outflows from the WR and O stars in the cluster. There are several other pieces of evidence as well which associate very high energy sources (MeV to TeV) to the surrounding star formation region of WR 121a \citep[see][]{2008AIPC.1085..372C, 2013ApJ...773...77A, 2016A&A...587A..93R, 2017MNRAS.468.2093D, 2018A&A...612A...1H, 2019A&A...627A..13B}.   

Our motivation to carry out the study of WR 121a is to explain the underlying mechanism of the X-ray emission from it. Many times speculations were made for this object to be a massive binary system but a systematic confirmation is required. If it is a colliding wind binary (CWB) then it should reveal some hints of variability in its long-term observations. This information will help us to classify this source among the category of other massive binaries. An observational study of this kind is necessary to constrain the stellar winds interaction features of as many objects as possible and to present some inputs to existing theoretical models that describe the physics of colliding winds to make more accurate predictions.  The studies carried out previously based on the similar approach have enriched us with valuable information \citep[e.g.][etc.]{2006MNRAS.371.1280D,2008MNRAS.390L..78O,2012MNRAS.422.1332Z,2014ApJ...788...84P,2015A&A...573A..43L,2015ApJ...799..124S,2019MNRAS.487.2624A}. Also, WR 121a lies in a rich complex of star-forming region towards the galactic center, therefore, it is important to know about the nature of massive stars present in that region to quantify their contribution towards different dynamical processes occurring there.

This paper is organized as follows. Section\,\ref{observ} presents the observations used and the data reduction methodology. The X-ray timing analysis is given in Section\,\ref{lc}.  In Section\,\ref{spectra}, we  present the X-ray spectral analysis of WR 121a. Our main results are discussed in Section\,\ref{disc} and Section\,\ref{conc} presents the conclusions.

\section{Observations and Data reduction.}\label{observ}

We have made use of archival X-ray data of WR 121a observed with \textit{Chandra} and \textit{XMM-Newton} for a total of 10 epochs from October 2004 to July 2016. The detailed log of observations is given in Table \ref{tab:log}. The procedure followed to process and reduce these data-sets has been mentioned in this section below.

\begin{deluxetable*}{c c c c c c c c c c c c}
\tablenum{1}
\tablecaption{Log of observations of WR 121a. \label{tab:log}}
\tablewidth{0pt}
\tablehead{
	\colhead{Sr.} 	& \colhead{Satellite} & \colhead{Obs. ID}  & \colhead{Obs. Date}  & \colhead{Start time}  & \colhead{Detector}  & \colhead{Livetime}    &\colhead{Source$^a$}     & \colhead{Offset$^b$}      & \colhead{Roll pointing}     \\ 
	\colhead{No.}     &  \colhead{}        & \colhead{}     & \colhead{}  & \colhead{(UT)}  &   \colhead{}  & \colhead{(ksec)}  &\colhead{counts}     & (\arcmin)   & \colhead{angle ($^{\circ}$)}}	
\startdata
            1 &   \textit{XMM-Newton}  & 0203850101  &   2004-10-22   &  05:48:23 &  PN       &  21.8 &  4303   & 0.136    & 258.74$^c$\\
               &                &                &           & & MOS1     &  25.3 &  2064   &          &          \\
               &                &                &           & & MOS2     &  25.3 &  1892   &          &          \\ 
            2 &  \textit{Chandra} & 9612       &   2008-05-27   &  16:58:20 &  ACIS-S       &   1.6 &  121    & 0.489    &  121.26   \\
            3 &  \textit{Chandra} & 18868       &   2016-06-13   &  16:27:05 &  ACIS-I       &  13.9 &  674    & 0.008    &  144.20   \\
            4 &   \textit{Chandra} & 17716       &   2016-06-14   &  03:33:28 &  ACIS-I      &  43.5 &  2370   & 0.008    &  144.20   \\
            5 &   \textit{Chandra} & 18870       &   2016-06-16   &  02:09:55 &  ACIS-I       &  32.7 &  1746   & 0.008    &  144.20   \\
            6 &   \textit{Chandra} &  18867       &   2016-06-16   &  19:28:17 &  ACIS-I      &  47.4 &  2429   & 0.008    &  144.20   \\
            7 &    \textit{Chandra} & 18869       &   2016-06-19   &  07:33:11 &  ACIS-I      &  39.5 &  2516   & 0.008    &  144.20   \\
            8 &   \textit{Chandra} &  18887       &   2016-07-28   &  09:05:55 &  ACIS-I      &  77.5 &  3425   & 0.008    &  235.20   \\
            9 &   \textit{Chandra} &  17717       &   2016-07-29   &  15:25:49 & ACIS-I      &  45.5 &  2559   & 0.008    &  235.20   \\
            10 &   \textit{Chandra} &  18888       &   2016-07-31   &  09:47:19 & ACIS-I      &  45.0 &  2332   & 0.008    &  235.20   \\
\enddata
\tablecomments{(a)  Background corrected net  source counts has been estimated in 0.5-10.0 keV energy range for \textit{XMM-Newton} while in 0.5-8.0 keV for \textit{Chandra}. \\
(b) Offset between WR 121a position and the telescope pointing.\\
(c) The position angle of the telescope has been mentioned in case of \textit{XMM-Newton} data-set. }
\end{deluxetable*}

\subsection{\textit{Chandra}}
W43$-$Main cluster was observed with \textit{Chandra} \citep{2000SPIE.4012....2W} for several times during June$-$July 2016 with different configurations of Advanced Chandra Imaging Spectrometer (ACIS)-I \citep{2003SPIE.4851...28G}. PI of these observations was Leisa K. Townsley. WR 121a was present almost at the aim-point of the detector for these data-sets. The exposure times of these observations were in the range of 13.9 to 77.5 ksec. The pointing Right Ascension and Declination of the telescope was almost the same for all of these observations but the satellite roll pointing angle was different for the data observed in June 2016 than those of July 2016 data-sets (see Table \ref{tab:log}). WR 121a was also observed by \textit{Chandra}$-$ACIS-S in May 2008 with relatively shorter exposure time and at a position with a larger offset from the detector aim-point (PI: Stephen Murray). 

\textit{Chandra} data sets were reduced using the standard reduction techniques\footnote{http://cxc.harvard.edu/ciao/threads/} with \textsc{ciao} version 4.11 and \textsc{caldb} version 4.8.2. The level$=$1 event files were reprocessed using the script \textsc{chandra\_repro}. In order to correct the arrival time of the X-ray photons, the barycentric correction was applied to the reprocessed, level$=$2 event files using the CIAO task \textsc{axbary}. Subsequently, the background light curves for each observation were produced and none was found to be affected significantly by high background rates.

\begin{figure*}
\gridline{\fig{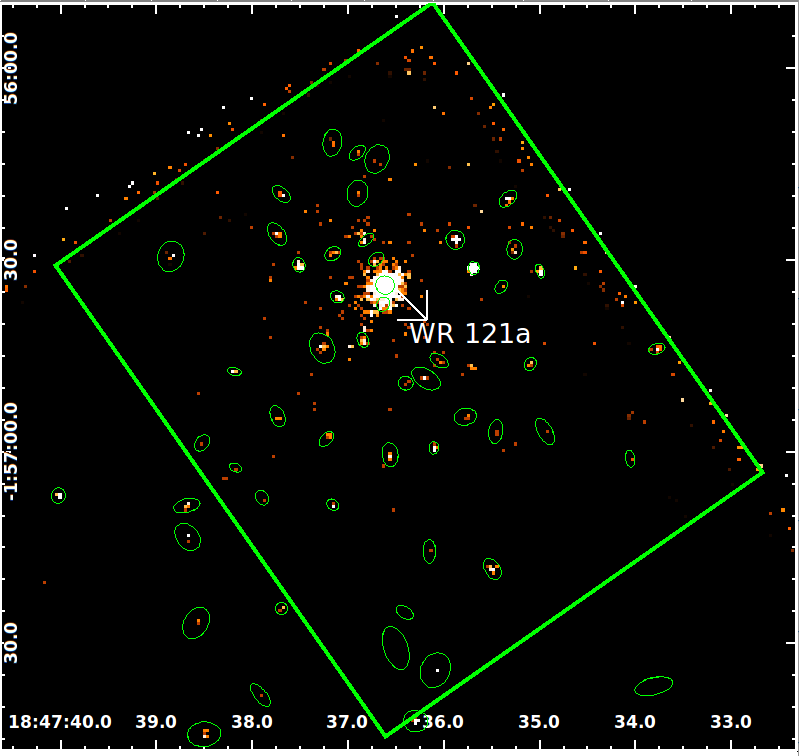}{\columnwidth}{}
          \fig{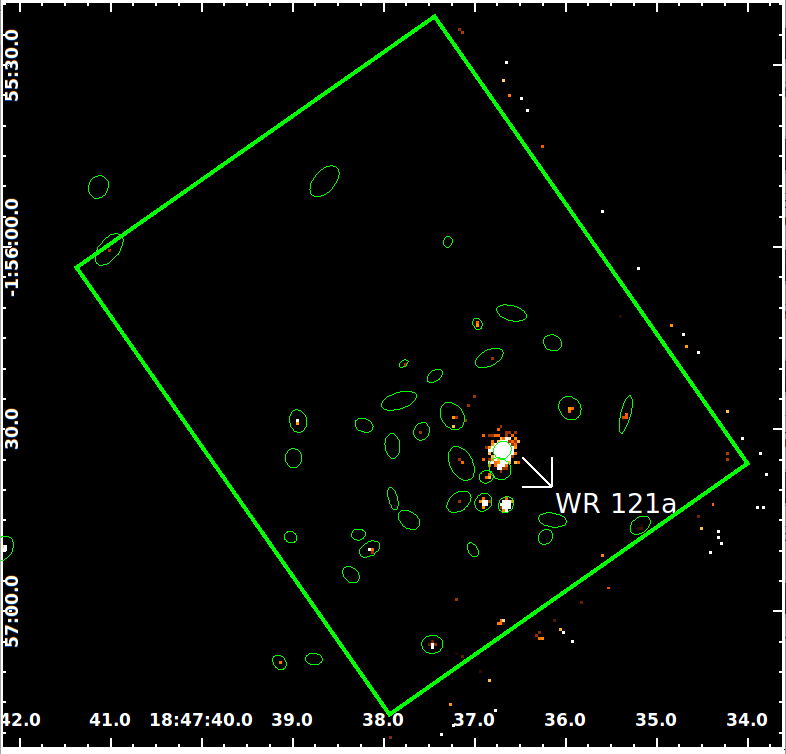}{\columnwidth}{}
          }
\caption{Merged and  exposure corrected  images of W43-Main from the observation IDs 18868, 17716, 18870, 18867, and 18869 (left) and 18887, 17717, and 18888 (right) of  \textit{Chandra}$-$ ACIS-I observations in the 0.5$-$7.0 keV energy band. The images are binned into $\sim$ 0.5\arcsec$\times$ 0.5\arcsec pixels (1 $\times$ 1 binning).   The 1.5\arcmin $\times$ 1.2\arcmin ~region of the W43-Main around its center has been marked by a rectangle. The X- and Y- axes correspond to RA (J2000) and DEC (J2000), respectively. The position of WR 121a as well as various other sources detected in the field has also been marked in the figures.
\label{fig:fig7}}
\end{figure*}

To determine the position of WR 121a accurately in the observed X-ray frames, we have merged different \textit{Chandra} observations. Another aim of merging different data-sets is to increase the signal to noise ratio and search for the binary companion in WR 121a, which was shown in ESO-VLT \textit{J}-band images by  \citet{2011A&A...532A..92L}, so that even if it is X-ray  faint it could be detected. The data-sets with different pointing roll angles were merged separately.  The first set is made of observation IDs with S.N. 3 to 7 (set 1) and another set 2 was made with S.N. 8 to 10  in Table \ref{tab:log}. Next, we ran \textsc{wavdetect} on each exposure independently to determine the position of various sources for matching between different exposures. The IDs 18867 and 18887 were chosen as references to align the different images of the set 1 and 2, respectively, due to their maximum exposure time among others in their respective sets. We then ran \textsc{wcs$\_$match} between all other exposures of each set with their respective reference image using sources returned from \textsc{wavdetect}. Subsequently, the aspect solutions as well as event files were updated with \textsc{wcs$\_$update} for each observation to align them with their respective reference image.

Finally, we produced two merged, exposure-corrected images for these two sets in 0.5$-$7.0 keV energy band and are shown in Figure \ref{fig:fig7}. The individual exposures of a set were first reprojected onto the frame of their respective reference image and then merged using CIAO task \textsc{reproject$\_$obs}. The merged images were then exposure corrected and binned into $\sim$ 0.5\arcsec$\times$ 0.5\arcsec pixels (1 $\times$ 1 binning) using the task \textsc{flux$\_$obs}. Further, the source detection algorithm \textsc{wavdetect}  was run onto the merged data and the position of the various sources detected has been shown with green ellipses in the Figure \ref{fig:fig7}.

WR 121a was detected at an average position  of R.A. (J2000) = 18:47:36.643 and Dec (J2000) = -01:56:33.77. The detected position of this X-ray source is closest to the position of WR 121a as given in the 2MASS All-Sky Catalog of Point Sources  \citep{2003tmc..book.....C}.  No nearby X-ray source was detected towards the North-East direction of the position of WR 121a in the merged images as shown by \citet{2011A&A...532A..92L} in the \textit{J}-band image.  The co-ordinates (R.A., Dec.) of W43 \#1a and \#1b as provided by \cite{2011A&A...532A..92L} are 18:47:36.691, -01:56:33.06 and 18:47:36.653, -01:56:33.22, respectively. The detected position of WR 121a lies closer to W43 \#1b toward its south.   Further, we have also  checked the possible structure of the X-ray source WR 121a.  A simulated \textit{Chandra} point spread function (PSF) image using the MARX\footnote{http://cxc.harvard.edu/ciao/threads/marx\_sim/} software tool was generated  following the CIAO science thread procedure as instructed. The CIAO task \textsc{srcextent} was used to  determine the structure of the source using this simulated PSF. The X-ray source centered at the position of WR 121a  was found  to be point-like with  90\% confidence level. The average position of the source detected nearest to WR 121a has the co-ordinates (R.A., Dec.) as 18:47:36.677, -01:56:36.72. It is at a distance of $\sim$3.0\arcsec ~towards the south of the detected position of WR 121a. The observed position of this closest source suggests that it is probably  W43 \#3 which is an O+O binary system \citep{2011A&A...532A..92L}. A detailed investigation is required to find the exact identity of this object.

  In order to generate the source light curves and spectra, a circular source region of radius 1.72$''$ was selected with a center position of  WR 121a. For background estimation, another source free circular region of radius 4.92$''$ was chosen, close to the source region, on the same ACIS-CCD chip for all the data-sets. The background subtracted X-ray light curves were extracted from these regions using the task \textsc{dmextract} in three energy bands \textit{viz.} 0.5-8.0 keV, 0.5-2.0 keV, and 2.0-8.0 keV. The spectra and the associated response matrices (ARF and RMF) were extracted using the task \textsc{specextract}. Individual spectra were grouped to have a minimum of 15 counts per energy bin.

\subsection{\textit{XMM-Newton}} 
WR 121a was observed with \textit{XMM-Newton} \citep{2001A&A...365L...1J} for a single epoch in  October 2004 (PI: David Helfand) with the European Photon Imaging Camera (EPIC) \citep{2001A&A...365L..18S,2001A&A...365L..27T}  instrument. EPIC consists of three CCDs namely MOS1, MOS2, and PN. The EPIC data were reduced with \textsc{SAS} version 17.0.0 using the latest calibration files. The SAS tasks \textsc{emchain} and \textsc{epchain} were used to produce processed event lists from the raw Observational Data Files (ODF) for the EPIC-MOS and EPIC-PN detectors, respectively. The SAS task \textsc{evselect} was used to extract the list of event files  with pattern 0 to 4 for PN and 0 to 12 for MOS data. The data were checked for pile-up using the task \textsc{epatplot} and it was found to be unaffected by this effect. Further, the data were also checked for the presence of high background intervals and it was found to be free from background flaring.

The X-ray light curves and spectra of WR 121a from EPIC instruments were generated from a circular region centered at the source position with a radius of 15\arcsec (equal to the half power diameter). In order to ascertain that no nearby sources contaminate the X-ray products generated for WR 121a, we have also identified sources present in the field using the task \textsc{edetect$\_$chain}. The source detected closest to the position of WR 121a lied outside the selected source region of 15\arcsec radius at a distance of $\sim$26\arcsec. However, \textit{Chandra} detected 3 sources using the observation ID 18869  (at the nearest phase) under the selected 15\arcsec  radius of WR 121a. These sources contribute only 3.5\% to the total count rate of WR 121a in 15\arcsec source radius of \textit{Chandra} observation.  Further,  we have converted count rates of all of these 3 \textit{Chandra} detected sources to the EPIC$-$PN of \textit{XMM-Newton} using the WebPIMMS\footnote{https://heasarc.gsfc.nasa.gov/cgi-bin/Tools/w3pimms/w3pimms.pl}. The corresponding count rates from EPIC$-$PN instrument for each source are equivalent or less than to the background count rate of the EPIC$-$PN for the present observation of 21.8 ks.  This could be a probable reason that these sources were not detected by \textit{XMM-Newton}. Under these circumstances, there is a low probability of contamination by neighboring sources to WR 121a in its 15\arcsec source radius in this \textit{XMM-Newton} observation. Nevertheless, we caution the readers to take the results obtained from \textit{XMM-Newton} with care against the possible contamination (if any) from neighboring sources.   Background estimation was also done from a circular region of the same size at source-free regions near the source on the same CCD. Light curves extracted from EPIC instrument were also corrected for good time intervals, dead time, exposure, point-spread function, and background using the \textsc{sas} task \textsc{epiclccorr}. The MOS-1 and MOS-2 light curves were added using \textsc{lcmath} of \textsc{heasoft} version 6.21. The task \textsc{evselect} was used to generate the source as well as the background spectra. To calibrate the flux and energy axes, the dedicated ARF and RMF response matrices were also calculated by the tasks \textsc{arfgen} and \textsc{rmfgen}, respectively. Back scaling of the extracted spectra was done using the task \textsc{backscale}. The EPIC spectra were grouped to have minimum 15 counts per spectral bin using \textsc{grppha}. Further temporal and spectral analyses were performed using \textsc{heasoft} version 6.21.

\section{X-ray timing analysis}\label{lc}
\begin{figure*}
\gridline{\fig{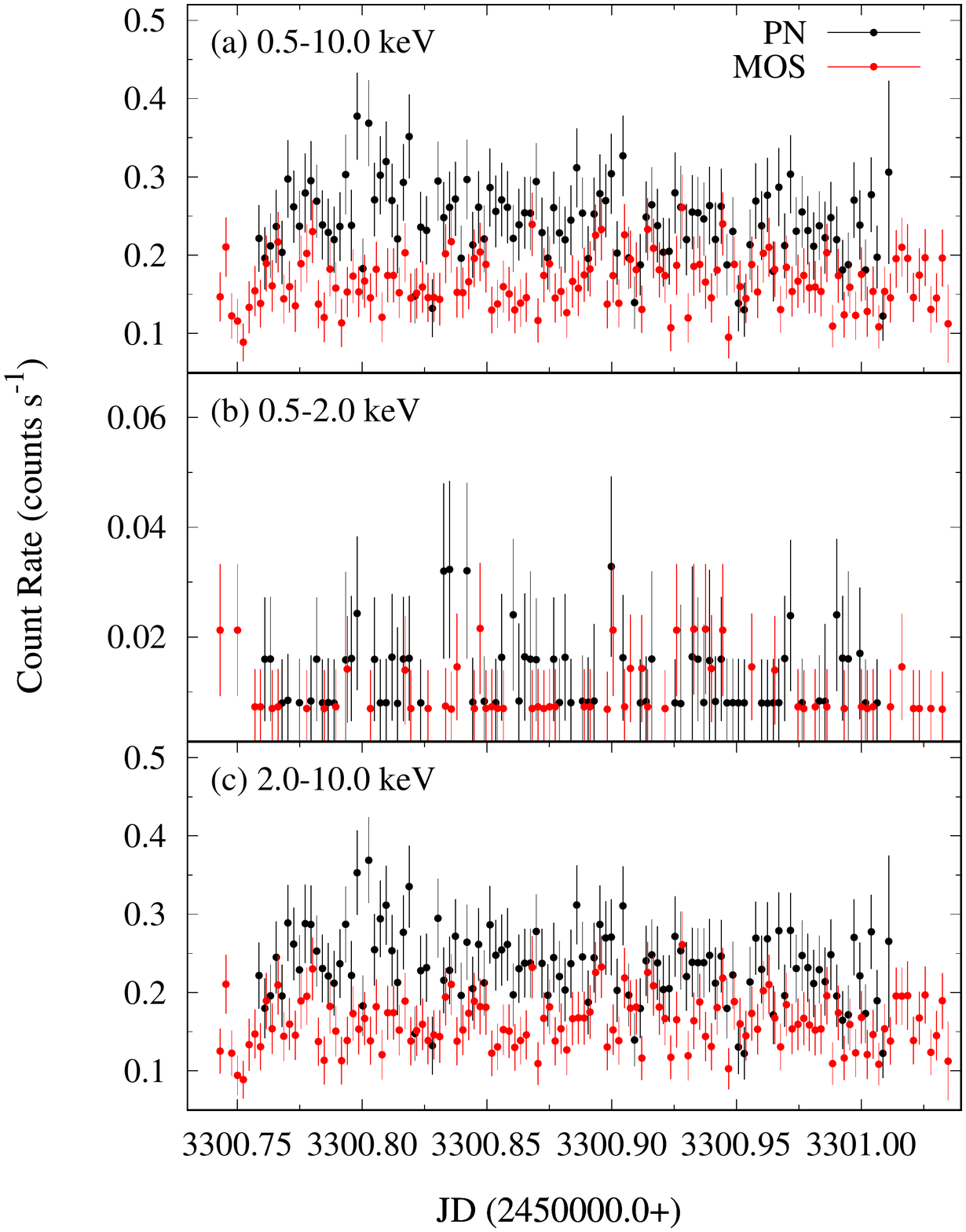}{0.5\textwidth}{(a)}\label{fig:fig1a}
          \fig{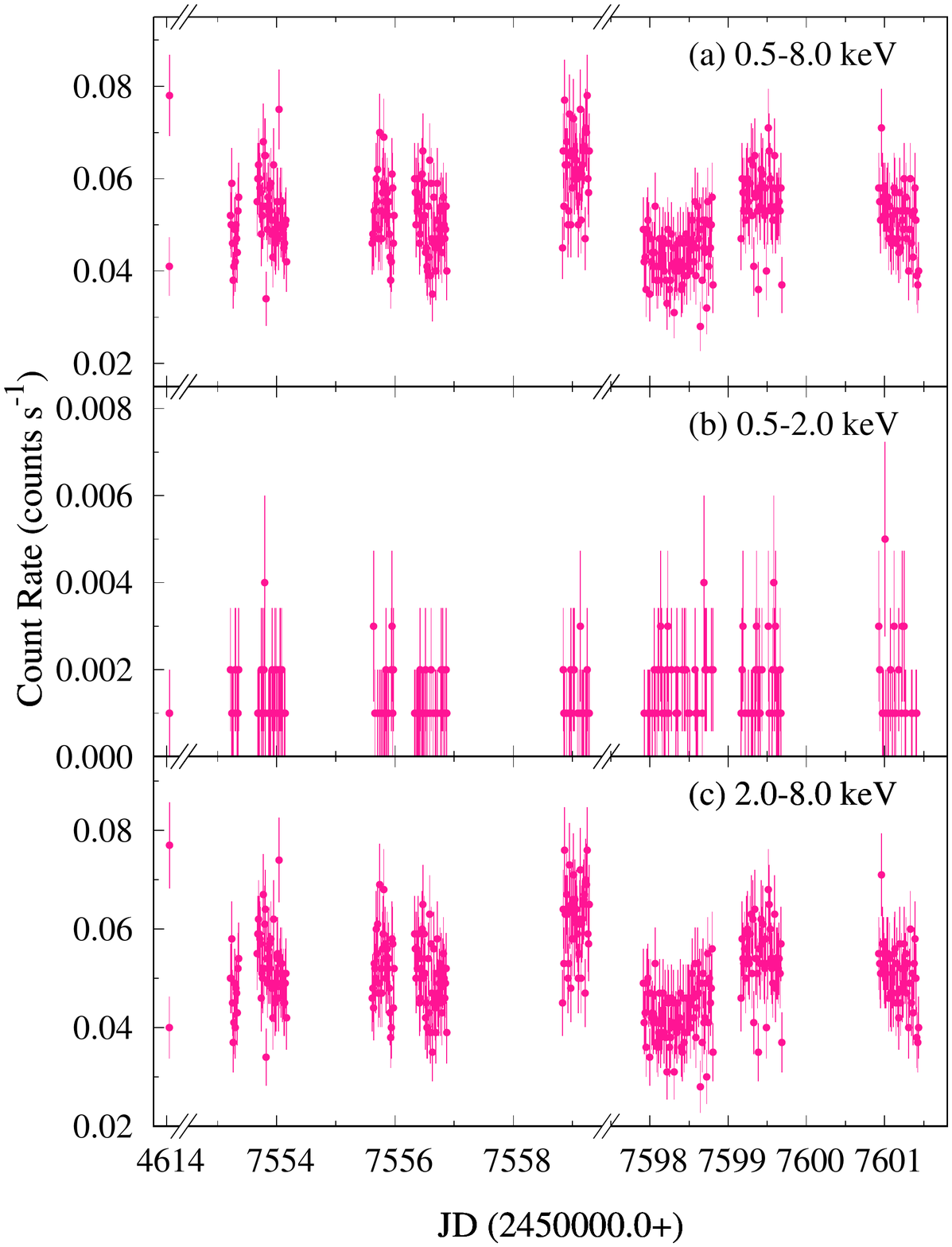}{0.5\textwidth}{(b)}\label{fig:fig1b} 
          }
\caption{X-ray light curves of WR 121a in different energy bands as observed by (a) \textit{XMM-Newton}-EPIC and (b) \textit{Chandra}-ACIS.
\label{fig:fig1}}
\end{figure*}

The background subtracted X-ray light curves as observed by \textit{XMM-Newton}$-$EPIC and \textit{Chandra}$-$ACIS were produced in the three energy bands and have been shown in Figure \ref{fig:fig1}. The \textit{XMM-Newton} light curves were binned with 200 sec while \textit{Chandra} light curves were binned with 1000 sec. The smaller number of count rate in the 0.5-2.0 keV energy band is indicative of the high absorption towards the source. However, the count rate in the other two energy bands follows a similar pattern. As shown in Figure \ref{fig:fig1}b,  a hint of variability is clearly seen in 0.5-8.0 and 2.0-8.0 keV energy bands.

\subsection{Search for variability}
 We have examined the collective \textit{Chandra}-ACIS light curve of WR 121a in the 0.5$-$8.0 keV energy range using the  $\chi^2$-test of variability  defined as

\begin{equation}
\chi^2 = \sum_{i=1}^{N}{\frac{(C_i-\bar{C})^2}{\sigma_i^2}}
\end{equation}

\noindent
where $\bar{C}$ is the average count rate, $C_{i}$ is the count rate of $i^{th}$ observation and $\sigma_i$ is the error corresponding to $C_{i}$. The value of $\chi^2$ obtained for the \textit{Chandra}$-$ACIS light curve in the 0.5-8.0 keV energy band is 1346 for 379 degrees of freedom. The $\chi^2$ statistic was compared against a critical value ($\chi_{\nu}^2$) for 0.1 \% significance level, obtained from $\chi^2$ probability distribution function. The $\chi^2$ value obtained is very large as compared to the $\chi_{\nu}^2$ of 470 for 379 degrees of freedom. This confirms that WR 121a was essentially variable with a 99.9 \% confidence level during these observations.  Light curves of the individual epoch of observations were also examined for the presence of variability using the $\chi^2$- test.  None of them was found to be significantly variable.

\begin{figure*}
  \centering
  \begin{minipage}{\columnwidth}
    \includegraphics[width=\columnwidth]{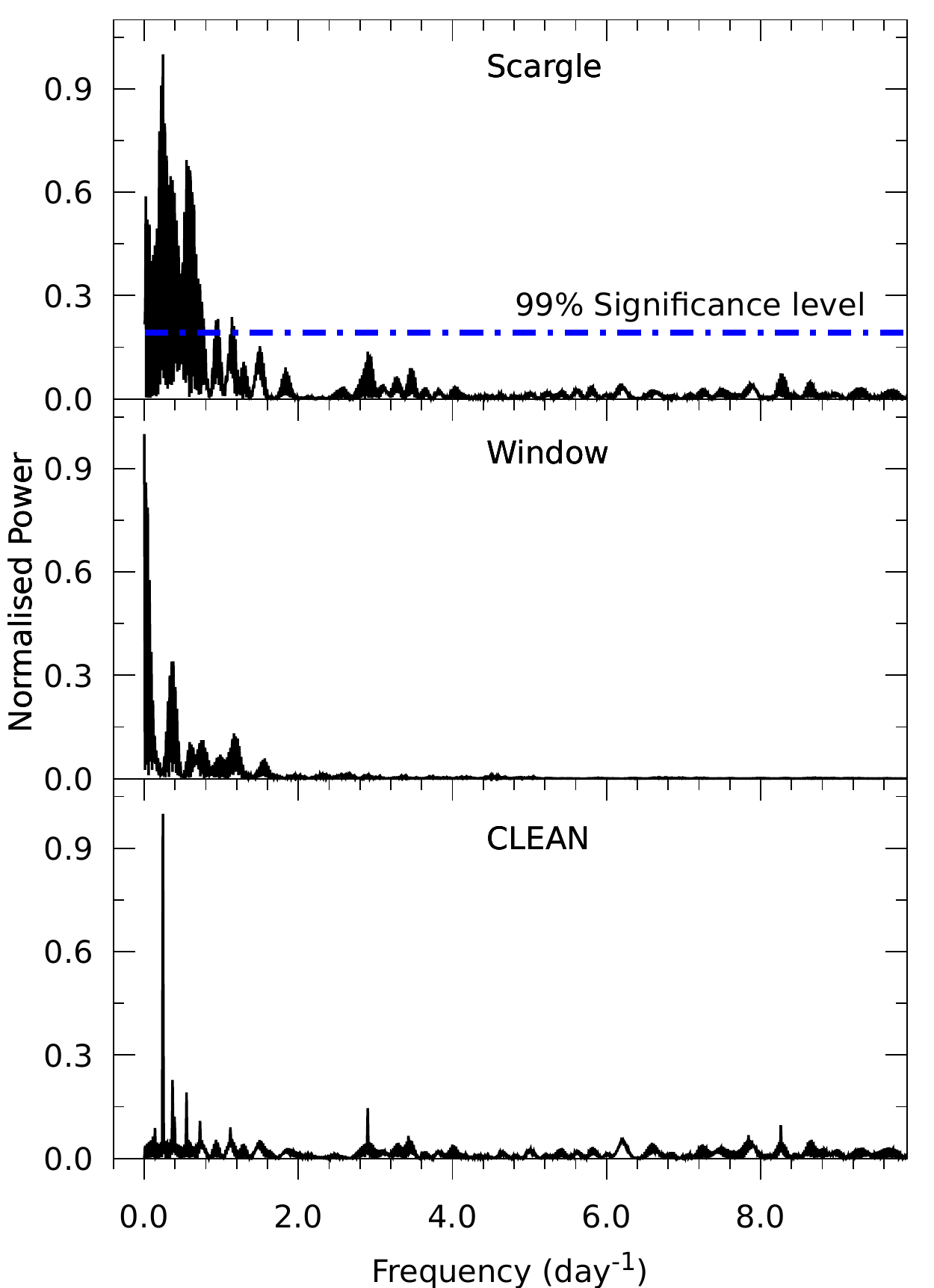}
    \caption{Lomb-Scargle (\textit{top}), data window (\textit{middle}), and CLEAN (\textit{bottom}) power spectra of WR 121a in 0.5-8.0 keV energy range using data from  \textit{Chandra}-ACIS. The dashed-dotted line in the top panel shows the false alarm probability corresponding to  99\% significance level.  \label{fig:fig2}}
  \end{minipage}
  \hfill
  \begin{minipage}{\columnwidth}
    \includegraphics[width=1.02\textwidth,trim={0.0cm 2.5cm 0.0cm 0.0cm}]{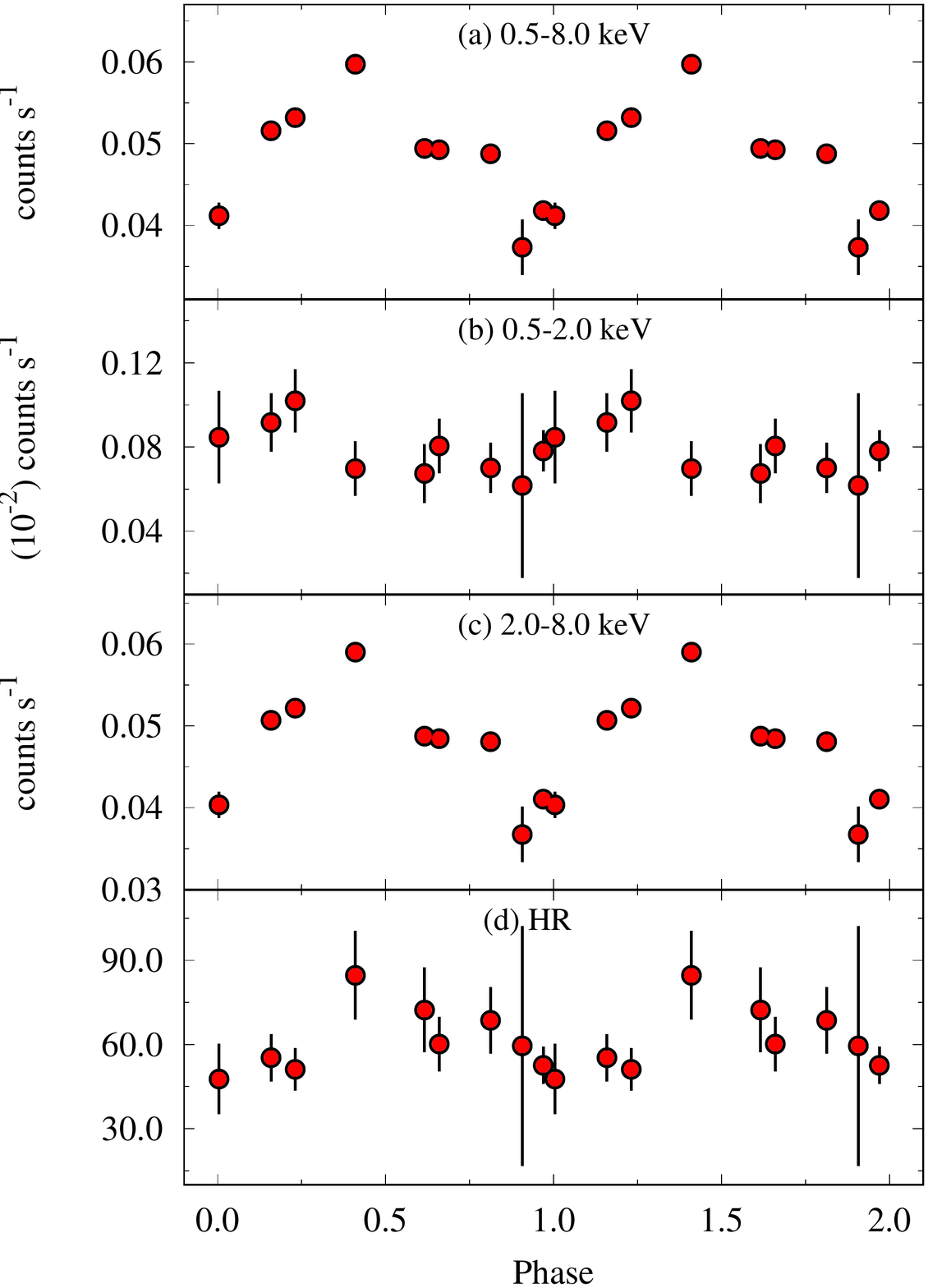}
    \caption{Folded X-ray light curves of WR 121a in different energy bands as observed by \textit{Chandra}$-$ACIS. \label{fig:fig3}}
\vspace{1.5cm}
  \end{minipage}
\end{figure*}

Further, we have performed a Fourier Transform (FT) of the \textit{Chandra}$-$ACIS light curve in the 0.5-8.0 keV energy band using the Lomb-Scargle periodogram \citep{1976Ap&SS..39..447L, 1982ApJ...263..835S, 1986ApJ...302..757H} to find any periodic signal. This is particularly effective in determining periodicities in those time series which are obtained over unequally spaced intervals of time. The top panel of Figure \ref{fig:fig2} shows the Lomb-Scargle power spectra of the \textit{Chandra} light curve in the 0.5-8.0 keV energy band, where a  peak power corresponding to a period of $4.1 \pm 0.1 $ days was noticed. We have also calculated a false alarm probability (see \citealt{1986ApJ...302..757H}) to check the significance of detected peaks. It is shown by the horizontal line in the top panel of Figure \ref{fig:fig2}.  Several other peaks were also present in the Lomb-Scargle power spectra, which may be due to the aliasing. To verify whether the identified peak in the periodograms is caused by the purely periodic signal or by the sampling, we have obtained the periodogram called the data window as shown in the middle panel of Figure \ref{fig:fig2}. It has been determined by setting all the data values as well as its variance to unity which were used to estimate the Lomb-Scargle periodogram of the light curve. Therefore, it simply depends upon how the sampling of the signal was performed and does not depend upon the signal itself. The identified peak in the  Lomb-Scargle power spectra did not fall under the window function indicating that the derived periodicity is real. We have also performed a periodogram analysis of the data using the  CLEAN algorithm \citep{1987AJ.....93..968R}. The CLEAN power spectrum was obtained by using a loop-gain of 0.1 and the number of iteration of 100.  Similar to the Lomb-Scargle periodogram, a peak corresponding to a period of $4.1 \pm 0.1$ d was also found in the CLEAN power spectra (see bottom panel of Figure \ref{fig:fig2}).

\subsection{Folded X-ray light curves}

X-ray light curves of WR 121a were folded using the ephemeris HJD$=$2457553.27+4.1E, where zero phase corresponds to a minimum X-ray count rate as observed for the Obs. ID 18868 and are shown in Figure \ref{fig:fig3}. 
The average count rate of an individual  observation was taken as a single data point in the folded light curve as light curve of individual  observation was non-variable. 
 In the energy band 0.5-2.0 keV, the average count rate belonging to the Obs. ID 9612 was found to be significant with 1.4$\sigma$ level, whereas average count rates for other Obs. IDs were significant by more than 3$\sigma$ level.   The light curves in the individual bands show phase-locked variability.  X-ray light curves in  0.5-8.0 keV and 2.0-8.0 keV energy bands show similar behavior.  In both of these energy bands, the count rate increased from the the  phase $\sim$ 0.0 to  a  phase $\sim$0.4 and then decreased up to the phase $\sim$1.0.  However, in the 0.5-2.0 keV energy band, initially, the count rates were increased from phase 0.0 to a phase $\sim$0.23. Then, it dropped in the later phase and became constant up to a phase $\sim$1.0 within a 1$\sigma$ level. The ratio of the maximum to the minimum count rates in 0.5$-$8.0 keV, 0.5$-$2.0 keV, and 2.0$-$8.0 keV energy bands was found to be  1.60$\pm$0.15, 1.51$\pm$0.38, and 1.61$\pm$0.15, respectively.  

The hardness ratio (HR), which is defined as the ratio of count rates in 2.0-8.0 keV and 0.5-2.0 keV energy bands, can reveal information about the spectral variations. The HR curve as shown in panel (d) of Figure \ref{fig:fig3} displays similar behavior to that of the light curve in the 2.0$-$8.0 keV energy band. The maximum value of the HR near the orbital phase $\sim$0.4 indicates a harder spectrum relative to other orbital phases.

\section{X-ray spectral analysis}\label{spectra}

\begin{figure*}
  \centering
  \begin{minipage}{\columnwidth}
    \includegraphics[width=1.2\columnwidth,trim={0.0cm 0.0cm 0.0cm 0.0cm}]{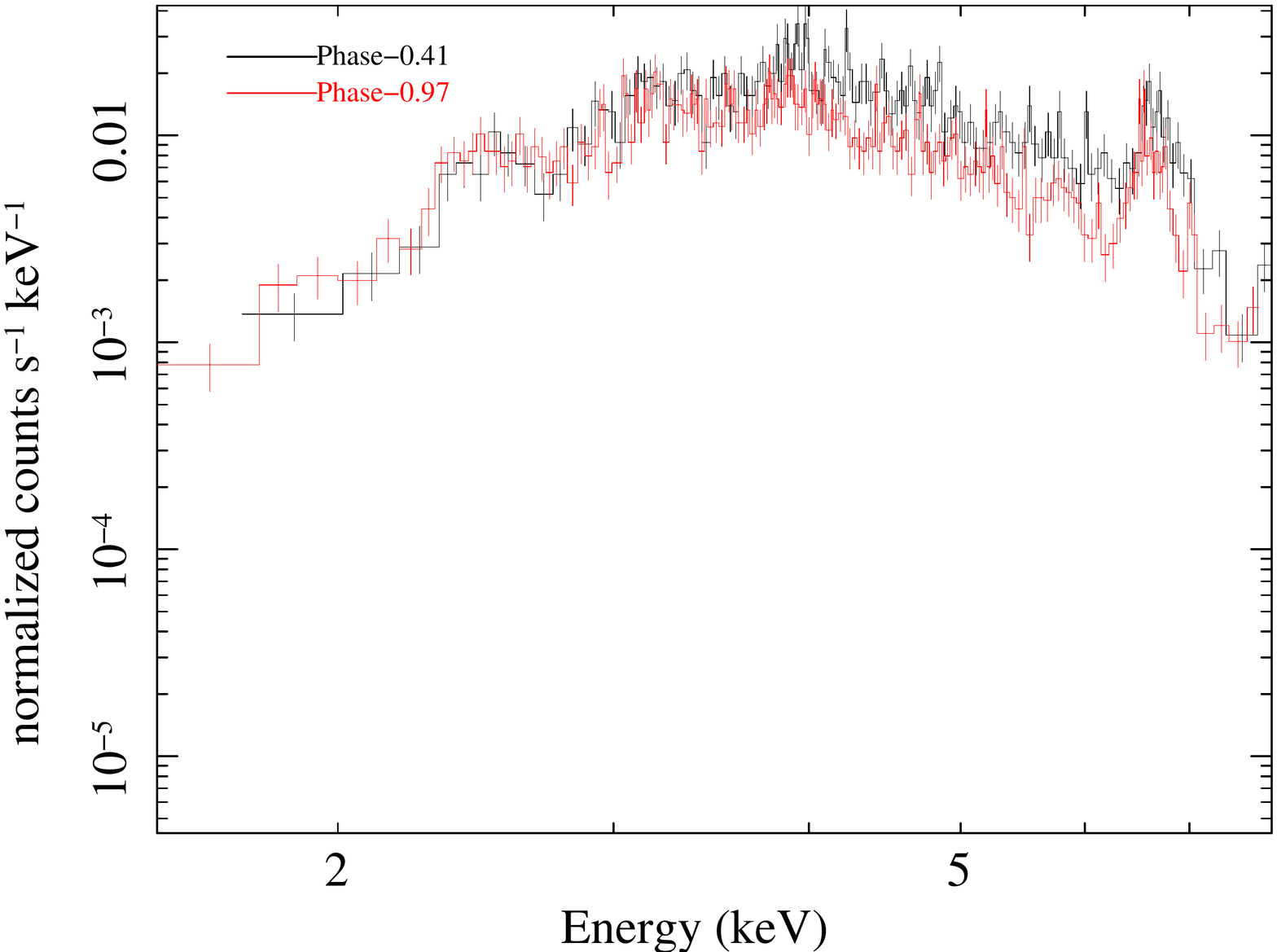}
    \caption{ Comparison of \textit{Chandra}$-$ACIS spectra of WR 121a  at two different orbital phases. Black points correspond to an orbital phase $\sim$ 0.41 and red points refer to a phase $\sim$ 0.97. \label{fig:fig4}}
    \vspace{1.1cm}
  \end{minipage}
  \hfill
  \begin{minipage}{\columnwidth}
    \includegraphics[width=1.2\textwidth,trim={0.0cm 0.0cm 0.0cm -3.0cm}]{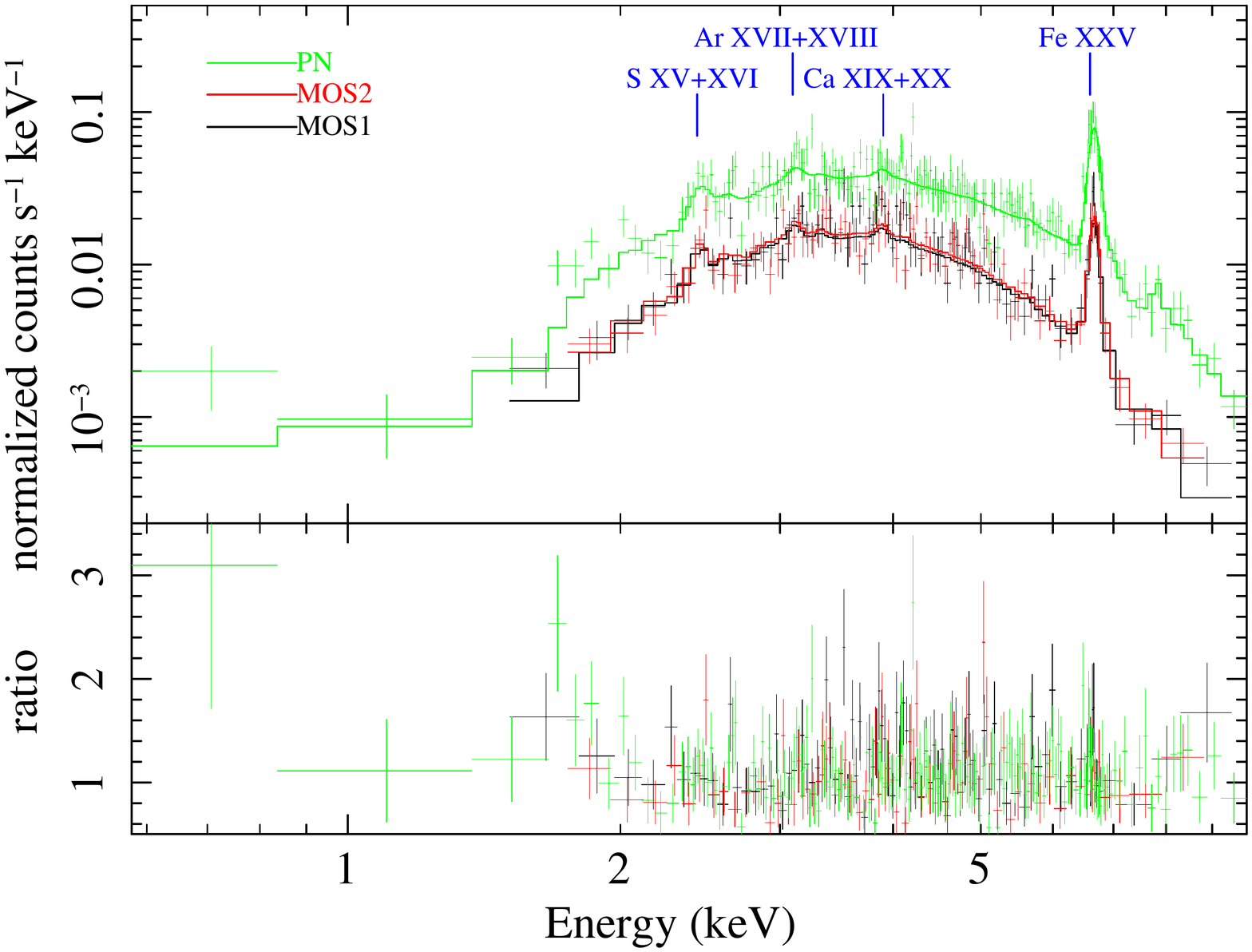}
    \caption{ X-ray spectra of WR 121a as observed from MOS and PN detectors onboard \textit{XMM-Newton} along with the best fit folded two-temperature plasma emission model (2T \textsc{apec}). The lower panel shows the residual in terms of the ratio of the data and model. \label{fig:fig5}}
\vspace{1.5cm}
  \end{minipage}
\end{figure*}

The \textit{Chandra}$-$ACIS-I spectra of WR 121a at different orbital phases have been shown in Figure \ref{fig:fig4}. Several emission lines are visible in the X-ray spectrum of WR 121a. The two spectra in the Figure \ref{fig:fig4} display the difference in the source flux when the two components of the binary system were $\sim 0.5$ phase apart. The soft energy part is highly absorbed as predicted for this deeply embedded object. But the system appears to be slightly brighter around phase 0.41 than phase 0.97 while this difference vanishes below $\sim$4\, keV. The HR curve as shown in the bottom panel of Figure \ref{fig:fig3} also imparts a similar feature. 

The best fit model to fit the X-ray spectrum of WR 121a has been obtained using \textit{XMM-Newton} data since the X-ray spectra from EPIC instruments have the better photon statistics than those of \textit{Chandra} data (see Table \ref{tab:log}). We have even attempted to fit combined spectra from all the 9 data-sets of \textit{Chandra} to obtain the best-fit parameters. But it returned some bizarre values after spectral fitting. Therefore, we began with the fitting of MOS and PN spectra of WR 121a jointly in 0.5$-$10.0 keV energy range using the models of the Astrophysical Plasma Emission Code (\textsc{apec}, \citealt{2001ApJ...556L..91S}) implemented in the X-ray spectral fitting package \textsc{xspec} \citep{1996ASPC..101...17A} version 12.9.1. We could not derive the abundances of individual elements using the  \textit{XMM-Newton} spectra of WR 121a by using the model ``\textit{vapec}'' in \textsc{xspec} as the best fit model returned to unphysical values of abundances. Therefore, we decided to fit the spectrum with a model in the form of \textit{$phabs(ism)*phabs(local) (apec+apec)$}. The component \textit{phabs(ism)} and  \textit{phabs(local)} were used to model the interstellar absorption and the local wind absorption effects, respectively, with elemental abundances according to \citet{1989GeCoA..53..197A}. The parameter N$_H^{ISM}$ corresponding to the model component ``\textit{$phabs(ism)$}'' was frozen at the value of 6.5 $\times$ 10$^{22}$ cm$^{-2}$ \citep{2011A&A...532A..92L}. The abundances (\textit{Z}) were tied for both components of  ``\textit{apec}'' model. All other parameters along with Z  were kept free in the fitting. A $\chi^{2}$ statistics was used for the spectral fitting. The best fit model on \textit{XMM-Newton}$-$EPIC spectra returned the values of two temperatures (\textit{i.e.} \textit{kT}$_{1}$ and \textit{kT}$_{2}$) as 0.98 $\pm$ 0.34  keV and 3.55 $\pm$ 0.69 keV, respectively. However, \textit{Z} was found to be   0.8 $\pm$ 0.1 times of solar photospheric abundances. The blending of various emission lines in the low resolution X-ray spectrum of WR 121a as well as non-detection of few lines below 2.5 keV due to heavy absorption leads to sub-solar global abundances in the spectral fitting. Therefore, high resolution grating spectra are needed for precise abundance determinations. The X-ray spectra of WR 121a as observed by MOS and PN along with the best fit model has been shown in Figure \ref{fig:fig5}. Various spectral lines along with the continuum have been fitted properly with a reduced $\chi^{2}$of 0.99 for 351 degrees of freedom. The strong emission line has been detected between 6 and 7 keV. The centroid of the line has been estimated by fitting a Gaussian profile and found to be at  6.66$\pm$0.01 keV with an equivalent width of 1.38$\pm$0.15 keV, which is similar to that found by \cite{2011ApJ...727..105A}. This line has been identified as Fe \textsc{xxv} and remains unresolved in the \textit{XMM-Newton} spectrum of WR 121a since \textit{XMM-Newton}  possess energy resolution of  $\sim$135$-$140 eV (FWHM) around 7 keV.

\begin{figure}
\centering
\includegraphics[width=\columnwidth]{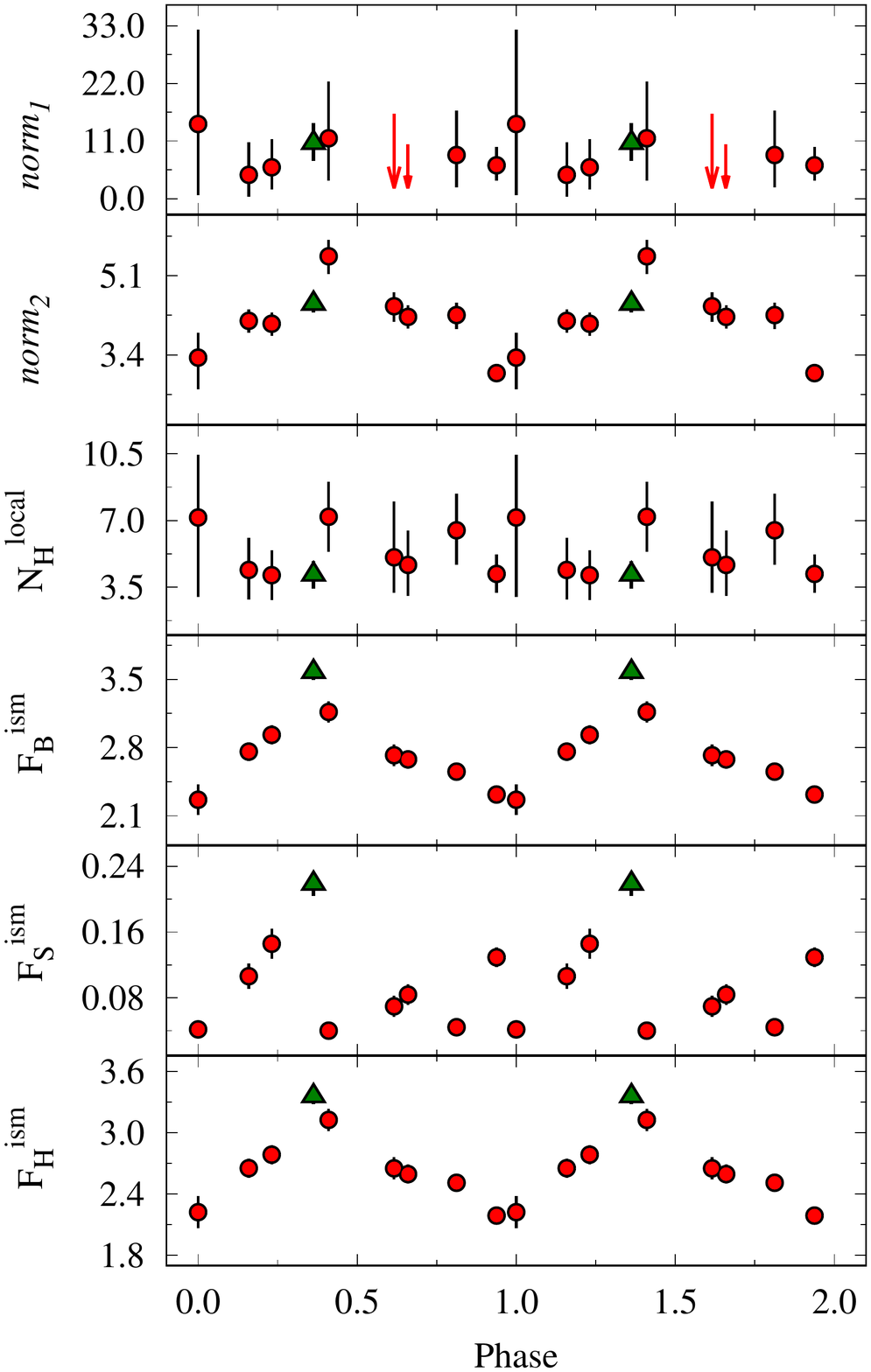}
\vspace{-1.0cm}
\caption{Spectral parameters as a function of the orbital phase of WR 121a obtained from \textit{Chandra}$-$ACIS (red filled circles) and \textit{XMM-Newton}$-$EPIC data (green filled triangles). Here, $norm_{1}$ and $norm_{2}$ are in the units of 10$^{-3}$ cm$^{-5}$, N$_{H}^{local}$ in  10$^{22}$ cm$^{-2}$, and F$_{B}^{ism}$, F$_{S}^{ism}$ as well as F$_{H}^{ism}$ are in the units of 10$^{-12}$ erg cm$^{-2}$ s$^{-1}$.  The downward arrows in the top panel mark the upper limits of $norm_{1}$ for two \textit{Chandra} Obs. IDs (see Table 2). \label{fig:fig6}}
\end{figure}

While fitting the \textit{Chandra} spectrum from individual observations, initially we have followed the same procedure  as applied to fit the EPIC spectra. But the temperature of both components over all phases were found to be consistent.  Thus,  \textit{Chandra} spectra of WR 121a  at different epochs of observations were fitted by keeping fixed the  parameters N$_H^{ISM}$, \textit{kT}$_{1}$, \textit{kT}$_{2}$,  and \textit{Z} to the best fit values obtained from the EPIC spectra.  Whereas the local equivalent H-column density ($N_H^{local}$) along with the normalization constants of both  temperature components ($norm_{1}$ and $norm_{2}$) were free parameters. Each spectrum from \textit{Chandra}-ACIS was fitted individually except one from Obs. ID 9612 due to its poor photon counts. Therefore, it was fitted jointly with the spectrum of Obs. ID 18887 which lies at its closest phase.  The ISM corrected X-ray fluxes of WR 121a in broad (0.3$-$10.0 keV, $F_{B}^{ism}$), soft (0.3$-$2.0 keV, $F_{S}^{ism}$), and hard (2.0$-$10.0 keV, $F_{H}^{ism}$) energy bands were derived using the model \textsc{cflux} in \textsc{xspec}. The intrinsic X-ray flux (\textit{i.e.} corrected for both the galactic as well as local absorption) has also been estimated in broad ($F_{B}^{int}$) energy band. The values of the best-fit parameters are given in Table 2. The variation of spectral parameters with orbital phase has been shown in Figure \ref{fig:fig6}, where the red filled circles and green filled triangles correspond to \textit{Chandra}$-$ACIS  and \textit{XMM-Newton}$-$EPIC data, respectively.  The parameters $norm_{1}$ and $N_H^{local}$ were found to vary along with the orbital phase within 1$\sigma$ level.  Whereas the parameters $norm_2$, $F_{B}^{ism}$ and $F_{H}^{ism}$ were varied in a similar fashion. A secular variation in these parameters was noticed, they increased from phase 0.0 to a phase near $\sim 0.4$ and then decreased up to a  phase 1.0. However, the variation in  $F_{S}^{ism}$ was found different than that seen in $F_{H}^{ism}$. At first, $F_{S}^{ism}$ increased from phase 0.0 to a phase near 0.36 then suddenly dropped to a phase 0.41, after this phase $F_{S}^{ism}$  increased again to a phase $\sim 0.6$ then showed a decreasing trend afterward.

\begin{deluxetable*}{c c c c c c c c c c}
\tablenum{2}
\tablecaption{Best fit parameters obtained from \textit{Chandra}$-$ACIS and \textit{XMM-Newton}$-$EPIC spectral fitting of WR 121a. \label{tab:spec_par}}
\tablewidth{0pt}
\tablehead{
\colhead{Obs.} & \colhead{Orbital}    & \colhead{\rm $norm_1$}  & \colhead{\rm $norm_2$}   & \colhead{\rm $N_{H}^{local}$}  & \colhead{\rm $F^{ism}_{B}$} & \colhead{\rm $F^{ism}_{S}$} & \colhead{\rm $F^{ism}_{H}$} & \colhead{\rm $F^{int}_{B}$} & \colhead{\rm $\chi^{2}_{\nu} (dof)$} \\
\cline{3-4}    \cline{6-9} 		 		
\colhead{ID}  &   \colhead{phase}   & \multicolumn2c{(\rm $10^{-3}$ cm$^{-5}$)} & \colhead{(\rm $10^{22}$ cm$^{-2}$)}  & \multicolumn4c{(\rm $10^{-12}$ erg cm$^{-2}$ s$^{-1}$)} &
}
\startdata
		0203850101  & 0.36 &  10.60$_{-3.44}^{+3.86}$   & 4.49$_{-0.18}^{+0.18}$ & 4.15$_{-0.73}^{+0.73}$ & 3.57$_{-0.08}^{+0.08}$ &  0.22$_{-0.01}^{+0.01}$ &  3.35$_{-0.07}^{+0.07}$ &32.74$_{-2.5}^{+2.5}$& 0.99 (351)  \\
     18868  & 0.00 &  14.26$_{-13.66}^{+17.99}$ & 3.34$_{-0.68}^{+0.54}$ & 7.17$_{-4.18}^{+3.29}$ & 2.26$_{-0.16}^{+0.16}$ &  0.04$_{-0.01}^{+0.01}$ &  2.22$_{-0.16}^{+0.16}$ & 39.53$_{-8.77}^{+8.74}$& 1.61 ( 36)  \\
     17716  & 0.16 &   4.54$_{ -4.20}^{+6.23}$  & 4.13$_{-0.26}^{+0.25}$ & 4.41$_{-1.54}^{+1.67}$ & 2.76$_{-0.09}^{+0.09}$ &  0.11$_{-0.02}^{+0.02}$ &  2.65$_{-0.09}^{+0.09}$ &  17.62$_{-3.25}^{+3.24}$ &1.25 (120)  \\
     18870  & 0.62 &  16.22$^a$       & 4.44$_{-0.34}^{+0.30}$ & 5.07$_{-1.86}^{+2.94}$ & 2.72$_{-0.11}^{+0.11}$ &  0.07$_{-0.01}^{+0.01}$ &  2.65$_{-0.11}^{+0.11}$ &17.10$_{-4.08}^{+4.08}$ & 0.99 ( 93)  \\
     18867  & 0.81 &   8.32$_{-6.14}^{+8.55}$   & 4.25$_{-0.29}^{+0.27}$ & 6.49$_{-1.82}^{+1.91}$ & 2.55$_{-0.09}^{+0.09}$ &  0.04$_{-0.01}^{+0.01}$ &  2.51$_{-0.09}^{+0.09}$ &26.89$_{-4.25}^{+4.25}$  &1.16 (121)  \\
     18869  & 0.41 &  11.53$_{-8.12}^{+10.81}$  & 5.52$_{-0.38}^{+0.35}$ & 7.19$_{-1.83}^{+1.84}$ & 3.16$_{-0.11}^{+0.11}$ &  0.04$_{-0.01}^{+0.01}$ &  3.12$_{-0.11}^{+0.11}$ &36.66$_{-5.76}^{+5.75}$  &1.03 (127)  \\
     17717  & 0.23 &   5.97$_{-4.26}^{+5.36}$   & 4.07$_{-0.26}^{+0.25}$ & 4.13$_{-1.31}^{+1.29}$ & 2.93$_{-0.09}^{+0.10}$ &  0.15$_{-0.02}^{+0.02}$ &  2.79$_{-0.09}^{+0.09}$ &20.97$_{-3.19}^{+3.18}$ & 0.99 (130)  \\
     18888  & 0.66 &  10.42$^a$                 & 4.21$_{-0.25}^{+0.25}$ & 4.68$_{-1.63}^{+1.81}$ & 2.68$_{-0.09}^{+0.09}$ &  0.08$_{-0.01}^{+0.01}$ &  2.59$_{-0.09}^{+0.09}$ &16.27$_{-3.25}^{+3.25}$ & 1.00 (119)  \\	
     18887  & 0.97   & \multirow{2}{*}{$6.37_{-2.96}^{+3.51}$} &   \multirow{2}{*}{$3.01_{-0.18}^{+0.17}$}   & \multirow{2}{*}{4.20$_{-0.99}^{+1.01}$} & \multirow{2}{*}{$2.32_{-0.07}^{+0.07}$} & \multirow{2}{*}{$0.13_{-0.01}^{+0.01}$} &  \multirow{2}{*}{$2.19_{-0.06}^{+0.06}$}  &\multirow{2}{*}{$20.19_{-2.26}^{+2.25}$} &\multirow{2}{*}{1.25 (175)}  \\
     9612  & 0.91  &  &       &   &  & &   &  &  \\  
\enddata   
\tablecomments{Fit parameters are based on a two-temperature \textsc{apec} model with two absorption components  $N_{H}^{ISM}$ (for galactic absorption) and  $N_{H}^{local}$ (for local wind absorption). The fitted model has the form \textit{$phabs(ism)*phabs(local)*(apec+apec)$} with $N_{H}^{ISM}=6.5 \times 10^{22}$ $cm^{-2}$, the two temperatures $kT_{1}$ = $0.98 \pm 0.34 $ keV and  $kT_{2}$ =  $3.55 \pm 0.69 $ keV, and abundances $Z$  = $0.8$ fixed to values derived from fitting of \textit{XMM-Newton}$-$EPIC spectra of WR 121a. $norm_{1}$ and $norm_{2}$ are the normalization constants for two temperature components. $F_{B}^{ism}$,  $F_{S}^{ism}$, and $F_{H}^{ism}$ are the ISM corrected X-ray fluxes of WR 121a in broad, soft and hard energy bands, respectively.   $F_{B}^{int}$ is the intrinsic X-ray flux corrected for both $N_{H}^{ISM}$as well as  $N_{H}^{local}$in broad energy band.  $\chi_{\nu}^{2}$ is the reduced $\chi^2$  and  \textit{dof} is degrees of freedom.  Errors quoted on different parameters refer to 90 per cent confidence level. The spectra obtained from Obs. IDs 18887 and 9612 has been fitted jointly. \\ $^a$ Mentioned values correspond to the upper limits of the specified parameter.}	
\end{deluxetable*}

\section{Discussion}\label{disc}

A detailed investigation of WR 121a using a long-term  X-ray data from \textit{Chandra} and \textit{XMM-Newton} satellites have been carried out for the first time. The X-ray spectrum of WR 121a below 10.0 keV represents the typical characteristics of an X-ray emission from an optically thin thermal plasma at temperature $>$10$^6$\, K (see Figure \ref{fig:fig5}).  These features are expected to arise from plasma heated in the wind collision region (WCR) of a  colliding wind massive binary where supersonic winds from the two massive stars produce shock-heated gas \citep{1992ApJ...386..265S}. This results in the production of  thermal hard X-ray emission. \citet{2011ApJ...727..105A} fitted \textit{XMM-Newton} spectra of WR 121a with a simpler one-temperature \cite{1977ApJS...35..419R} model attributed to the presence of a likely thermal component to the X-ray emission. The values of temperature and abundances derived by them are consistent with our estimates. Analyzing the X-ray spectrum of WR 121a, \citet{2011ApJ...727..105A} has also suggested that it belongs to the category of CWBs. The current study verifies this fact by looking at the following details: 

\begin{enumerate}
\item The maximum ISM corrected X-ray luminosity (\textit{$L_{X}^{ism}$}) of WR 121a in 0.3-10.0 keV energy band is estimated to be 1.54$\times$ 10$^{34}$ erg s$^{-1}$ (corresponding to a distance of 6 kpc). While the ISM and local wind absorption corrected X-ray luminosity ($L_{X}^{int}$)  has been estimated to  1.70$\times$ 10$^{35}$ erg s$^{-1}$. This belongs to the typical X-ray luminosity range of $\sim$10$^{32}$ to 10$^{35}$ erg s$^{-1}$  for WR $+$ O binary systems \citep[e.g.][]{2012ASPC..465..301G}. The derived X-ray luminosity for WR 121a is also found to be more than the typical X-ray luminosity of several single WN stars \citep{2010AJ....139..825S,2012AJ....143..116S}.
\item The presence of a strong 6.7 keV Fe \textsc{xxv}  emission line further supports this fact since its generation requires the high temperature plasma (approximately 10$^{7}$ K ) which is present in the WCR  as pointed out by \citet{2003A&A...402..653R} .  
\item The present theoretical models about CWBs suggests that a substantial amount of X-rays originate from the collisions of dense and highly supersonic stellar winds of binary components \citep{1976SvA....20....2P,1976SvAL....2..138C}. However, the individual stars may also have a considerable intrinsic soft X-ray emission due to the relatively less stronger shocks that develop as an outcome of the line driven instabilities \citep{1982ApJ...255..286L,1988ApJ...335..914O}. Therefore, the fitting of X-ray spectra of bright WR $+$ O binaries requires at least two thermal components, which is also seen in the case of WR 121a (see section \ref{spectra}).  The two temperatures derived from the spectral fitting of WR 121a  are well within a range typically found for massive binaries, e.g. $\leq$1 keV for cool and 2.0-4.0 keV for hot temperature component \citep{1997A&A...322..878F,2016AdSpR..58..761R}.
\item Another important aspect of CWBs is the phase-locked variability of the flux and$/$or the hardness of the X-ray emission. This depends upon the amount of absorption suffered by the X-rays through the orbit. Since soft X-rays are more strongly absorbed, therefore, increasing absorption reduces observed flux but increases the hardness. Figs. \ref{fig:fig3}, \ref{fig:fig4} and \ref{fig:fig6} display clear signatures of variability along the orbit and hence supports CWB status of WR 121a.
\end{enumerate}

\begin{figure}
\centering
\includegraphics[width=\columnwidth]{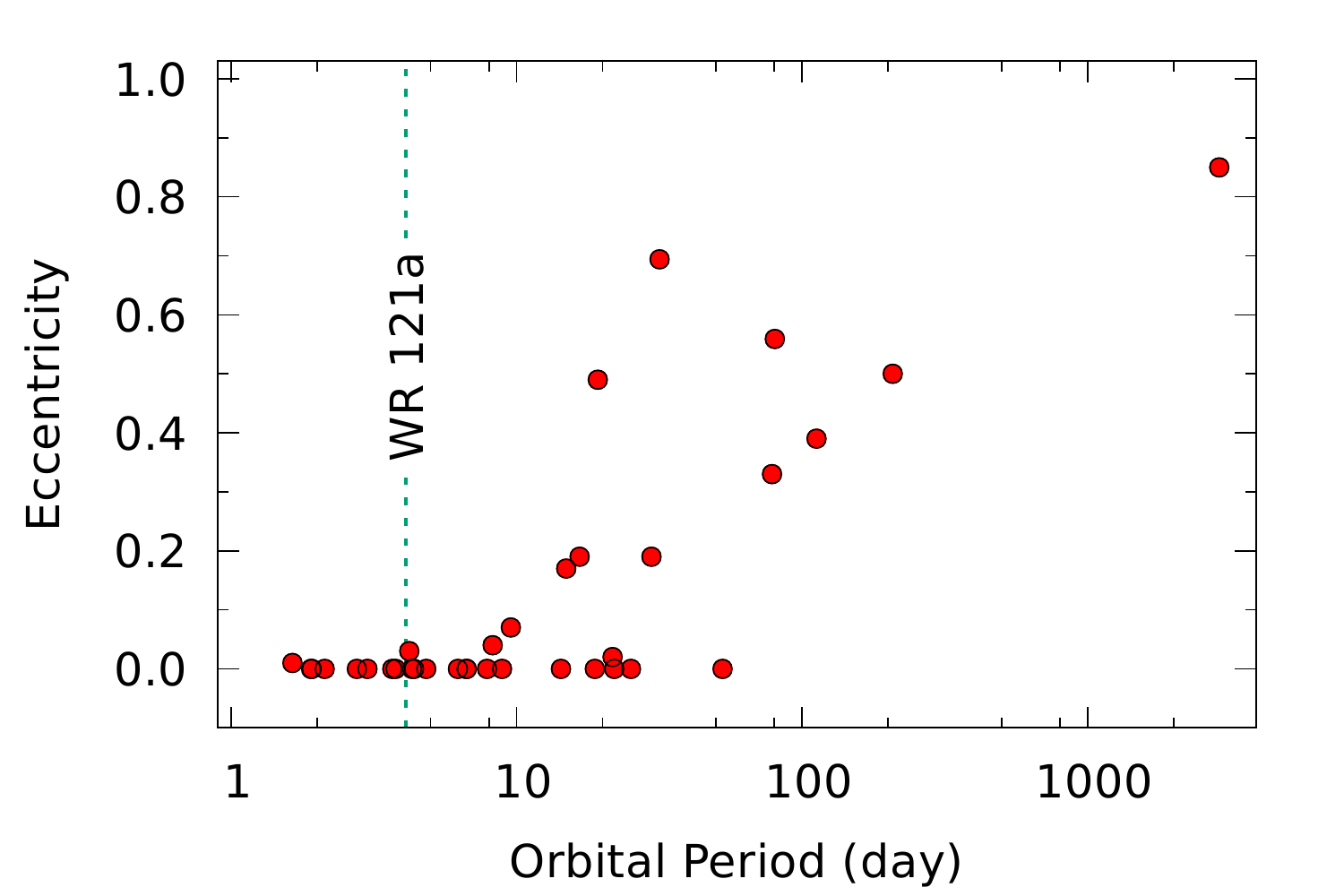}
\caption{Plot of eccentricity as a function of the orbital period of WR+O binaries. The location of WR 121a is shown by the dashed line. \label{fig:ecc}}
\end{figure}

The X-ray light curves of WR 121a in the 0.5-8.0 keV energy band shows periodicity corresponding to $4.1 \pm 0.1 $ d period. For the  WR+O binaries, \citet{2003ARep...47...38C} suggested that the orbits of WR+O binaries are circular if their orbital periods are less than $\sim$ 14 days. Figure \ref{fig:ecc} shows the plot between eccentricity and the orbital period of WR+O binaries. Location WR 121a is shown by a dashed line. From this figure, it appears that circularization occurs in the orbital period between 10 - 20 days. Thus, the orbit of  WR 121a can be considered as a circular. Further, it appears that the circular orbit implies a phase independent wind collision in between the stars. If we consider a circular orbit, the phase variation in X-ray flux could be due to the change in absorption by the variable wind density along the line of sight and$/$or occultation of the X-ray emitting region.  Since hard X-rays are mainly generated in the WCR, therefore, the minimum F$_{H}^{ism}$ at phase 0.0,  as shown in Figure \ref{fig:fig6}, could be due to the eclipsing of WCR behind the O-type companion.  We suggest that the WCR is eclipsed by O-star as the location of WCR  is close to the O-star  (as explained later in this section) and the size of O-star is larger than its companion WN star.  After the maximum flux at $\sim 0.36$,  F$_{H}^{ism}$ decreases gradually till the orbital phase $\sim$1.0.  This is also evident from the larger HR ratio as well as the harder \textit{Chandra} X-ray spectrum of WR 121a  near phase $\sim$ 0.4 as seen in Figures. \ref{fig:fig3}  and \ref{fig:fig4}, respectively.   The F$_{S}^{ism}$ also follows a similar trend as that of F$_{H}^{ism}$ during the orbital phase 0.0 to 0.36. But it shows instances of increased$/$decreased values around other orbital phases which is exactly opposite to the trend followed by $N_H^{local}$. Although it is difficult to anticipate anything from the variation of $N_H^{local}$, owing to its large error bars, still it presents hints about our line of sight passing through stellar winds of different densities. Since $N_H^{local}$ mostly affects the soft X-rays,  therefore, this might be giving rise to orbital variations in  F$_{S}^{ism}$. 
The estimation of $N_H^{local}$ with large uncertainty along with the sparse orbital sampling do not allow us to accurately determine the geometry of the binary components as well as of the WCR during the orbital period. Therefore, further observations of WR 121a with dense orbital sampling in different wavebands are urgently required to understand its binary nature more deeply. 

 \citet{2011A&A...532A..92L} has suggested that the separation between the two components of WR 121a is $\sim$0.598\arcsec. Using this separation, we have estimated the orbital period ($P$) of WR121a system by using the Kepler's third law as

\begin{equation}
P^{2}= \frac{4\pi^{2}a^{3}}{G(M+m)} 
\end{equation}

\noindent
where $a$ is the linear separation corresponding to 0.598\arcsec ~angular separation at a distance of 6 kpc, $G$ is the universal Gravitational Constant, $M$ and $m$ represent the masses of the primary and secondary star of the binary, respectively.  Because of the similarity of the $K$-band spectra of WR 121a and WR 131, as noted by \citet{1999AJ....117.1392B}, the value of $M$ is taken same as that for the WN7 star of WR 131 \textit{i.e.} equal to 44  M$_\odot$ \citep{2019A&A...625A..57H}.  Using the several values of $m$ for typical  O-type dwarf$/$giant$/$supergiant star from \citet{2012A&A...537A..37M}, the value of $P$ is found to be more than 2700 years.  Further,  it is very unlikely for the faraway companions to give rise to a strong wind-wind collision to emit significant X-rays. Till now, the massive binaries having an orbital period only up to a few years have been seen to harbor X-ray bright wind-wind collision  (e.g. 9 Sgr with an orbital period of 9.1 yr; \citealt{2016A&A...589A.121R}). Therefore, either the companion of WN7 star in WR 121a as described by \citet{2011A&A...532A..92L} may not be correct or this may be a triple system where X-ray modulations of 4.1 days observed in this study are originating from the shorter orbital components.  The radius of WN7 star has been estimated to be 7.85 R$_\odot$ by using its typical values of luminosity as $10^{5.54}$ L$_\odot$, temperature as 50 000 K \citep{2007ARA&A..45..177C}, and assuming a black body emission. Whereas, there is a large spread in the mass of WN7-type star in the range of 7$-$49 M$_\odot$ \citep{2019A&A...625A..57H}.  The mass, as well as the radius of O-type companion, has been noted to lie in a range of 16.46$-$66.89 M$_\odot$ and 7.39$-$23.11 R$_{\odot}$, respectively, by considering all the possible cases of dwarf$/$giant$/$supergiant star \citep{2012A&A...537A..37M}. Assuming a circular orbit scenario and the above mentioned mass estimations, the linear separation from center of WN7 star to the center of O-type star of WR 121a system is found to lie in the range of 30.90$-$52.75 R$_{\odot}$ which translates to an angular separation of 0.024$-$0.041 mas at a distance of 6 kpc. The WR 121a system appears to be either detached or semi-detached where O-type star is filling its Roche-lobe especially when it is in supergiant state.

To explore the nature of wind collision in WR 121a, the value of the cooling parameter ($\chi$) at the position of wind collision in between two stars has also been estimated. $\chi$ is defined as the ratio of the cooling time of the post-shock gas to the typical escape time from the shock region by \citet{1992ApJ...386..265S} as 

\begin{equation}
 \chi = \frac{v^{4} D}{\dot{M}}
\end{equation}

\noindent
 where $v$ is the pre-shock wind velocity in 1000 km\,s$^{-1}$ units, $D$ is the distance from the star to the shock in $10^{7}$ km and $\dot{M}$ is the mass loss rate in 10$^{-7}$ M$_{\odot}$\,yr$^{-1}$. If  $\chi\ll1$, then the wind collision is considered as radiative where the gas cools through rapid radiation emission. However,  $\chi \geq 1$ indicates an adiabatic cooling regime. Since the spectral features of WR 121a are similar to that of WR 131,  the  mass of WN7 star has been adopted as 44  M$_\odot$ for the rest of the analysis. However, a range of masses as mentioned above for the O-type star has been considered. Additionally, by assuming the typical values of mass-loss rate, radius, and terminal-wind velocity for a WN7 and O-type dwarf$/$giant$/$supergiant star \citep{2007ARA&A..45..177C,2019A&A...625A..57H,2012A&A...537A..37M}, the position of WCR was estimated and correspondingly the wind speeds of two stars was calculated at the WCR using the standard $\beta$-velocity law. It was seen that the stellar winds of both the components of WR 121a interact at sub-terminal speeds and the value of $\chi$ for a circular orbit of period 4.1 days is estimated to be $\ll$1 for both the winds of WN- and O-star at the position of WCR. This is as expected since the shocked-wind has not enough time to escape the WCR and cool-down via adiabatic expansion in a short period massive binaries, therefore, it cools rapidly through radiation emission. 

The plasma temperature derived from the spectral fitting of WR 121a can also be used to estimate the pre-shock velocity by using the relation from \citet{1990ApJ...362..267L}  as

\begin{equation}
kT_{sh, max} = 1.95 \mu v_{\perp,1000}^{2}  \quad \textrm{keV}  
\label{eqn:eqn4}
\end{equation}

\noindent
where $\mu$ is mean mass per particle in units of the proton's rest mass, which is 1.16 for a WN star \citep{2007MNRAS.378.1491S} and 0.62 for an O-type star \citep{2008ApJ...683.1052C}, and $v_{\perp,1000}$ is the wind velocity component perpendicular to the shock front in units of 1000 km s$^{-1}$. The average values of pre-shock velocities corresponding to the cool temperature (which is mostly  generated due to the radiation driven wind shock of individual component of binary)  for the WN and O components of WR 121a are found to be  $\sim$ 658 and $\sim$ 900 km s$^{-1}$, respectively. Whereas the pre-shock velocities in the WCR corresponding to the hot temperature are estimated to be as 1253 and 1714  km s$^{-1}$ for WN and O-type components of WR 121a system, respectively. These values are lower than the typical terminal velocity observed for the single WN and O-type stars. This may be happening because of two reasons. One is that a wide range of temperatures is covered by the shocked plasma in WCR. The hottest plasma is expected to lie near the stagnation point while the cooler plasma is present along the wings of the shock front. But the X-ray spectral fitting gives the average temperature prevailing in the WCR which will be less than the maximum shock temperature mentioned in the relation \ref{eqn:eqn4}.  Another probable reason could be that if the winds of individual binary stars have not reached terminal speeds before colliding, as it is likely for short-period binaries,  then the maximum shock temperature will further be decreased.

In addition to the presence of hydrodynamic shocks in the CWBs, some other effects have to be considered as well. These include alterations of the characteristics of pre-shock flow along with the microscopic phenomena prevailing in post-shock plasma. For massive binaries, the interaction of the stellar wind with the strong radiation fields of the component stars inhibits the acceleration of one star's wind and prohibits it to reach  v$_{\infty}$ \citep{1994MNRAS.269..226S}. On the other hand,  another outcome of the radiation pressure of the companion star called radiative braking was examined by \citet{1997ApJ...475..786G}. This effect is most favorable for those binary configurations in which the wind momenta of the component stars are highly imbalanced e.g. WR + O binaries.  For those systems, as the dominant star wind approaches close to the surface of companion star,  the radiative momentum flux of the companion star suddenly decelerates it. This process leads to significant modification of the bow shock geometry and  the wider opening angle of the shock cone. In some situations, it is impossible for the companion wind to sustain the momentum of the dominant wind. Radiative braking prevents the collision of incoming wind onto the surface of companion star for those scenarios. For example, consider the case of V444 Cyg (WR 139). It is a massive binary where the primary is a  WN5 star whereas secondary is an O6 star. They move around an orbit which is almost circular (eccentricity=0.03) with an orbital period of $\sim$4.2 d \citep{1950ApJ...112..266M,1994ApJ...422..810M}. Owing to the shorter period and strong stellar winds of its components, it was suggested that both the radiative braking and inhibition are required to explain the X-ray light curves of V444 Cyg by \citet{2015A&A...573A..43L}. They have also estimated the shock opening angle for wind collision in V444 Cyg and it was found to be approximately 75$^{\circ}$. The large value of shock opening angle confirmed that these two mechanisms play an important role in the stellar wind interaction of two components of this binary. Our analysis reveals that WR 121a is also similar to V444 Cyg in terms of its orbit. Their periods are almost same. We have also noticed that while calculating $\chi$, the position of the WCR was found to coincide with the position of the secondary star in the orbit for the O-type star later than O6.0, O5.5 and O5.0 for dwarfs, giants, and supergiant, respectively. This indicates that the WCR might be just collapsing onto the surface of the secondary star of WR 121a. The ratio of the companion star radius to the  binary separation is found to lie in the range of 0.22$-$0.51, where the WR wind is shocking onto the companion surface. To explore this scenario further, we have estimated the shock opening angle ($\theta$) for various combinations of WN7 and O-type star in a 4.1 days circular orbit according to    \citet{1993ApJ...402..271E} as

\begin{equation}
\theta \simeq 2.1\left(1-\frac{\eta^{\frac{2}{5}}}{4}\right)\eta^{\frac{1}{3}} \quad   \textrm{for} \quad \textrm{$10^{-4}\leq \eta \leq 1$}
\end{equation}
\noindent
where  $\eta$ =${{(\dot{M}v_{\infty}}})_{O}/{{(\dot{M}v_{\infty}}})_{WN}$ is the wind momentum ratio. It has been seen that $\eta$ varies from 0.018 to 0.73 for all combinations of WN7 primary  with different  O-type secondaries and correspondingly $\theta$ was found to lie in the range of $\sim$30$^{\circ}$ to 84$^{\circ}$.  The larger value of  $\eta$ and hence $\theta$ observed for early O-type companions indicates that the radiative braking is playing an important role in the WCR of WR 121a as early O-type stars carry sufficient wind momentum to reduce the effect of the WN wind momentum. However, as the secondary star moves to the later O-type stars, the decreasing shock opening angle points towards the scenario of either no or low radiative braking and hence WN star wind overpowers the O-star wind and the shock forms at or very near the O-star surface as inferred from $\chi$ calculations also.      

As mentioned in the beginning of this section, WR 121a lies towards the high end of the  \textit{L$_{X}$}  range observed for massive binaries. Therefore, it is one of the brightest WR$+$O binary in X-rays.  It was noted by \citet{2012ASPC..465..301G} that generally all of the binary systems with WR stars having supergiant early O-type secondaries have  \textit{L$_{X}$}  $>$ 10$^{33}$ erg s$^{-1}$.  However, the systems with supergiant late O-type star as a companion of WR star have $L_{X}^{int}$ $\sim$ 10$^{33}$  erg s$^{-1}$ e.g. CQ Cep (WN6$+$O9II-Ib; \citealt{2015ApJ...799..124S}) and WR 133 (WN5$+$O9I; \citealt{2005MNRAS.361..679O}). Since $L_{X}^{int}$ of $\sim$ 10$^{35}$  erg s$^{-1}$ for WR 121a is much higher than the WR systems with late-type supergiant companion, therefore, we anticipate that  WR 121a system may have an early O-type supergiant  companion with  WN7 primary. If it is so, then the radiative braking, as well as inhibition, would certainly be playing an important role in the WCR of WR 121a system. The denser wind of WN7 star may hide the spectral features specific to an  O-type companion as the bolometric luminosity of a O-type star  is almost similar to that of WN7 star  \citep{2007ARA&A..45..177C,2012A&A...537A..37M}. The \textit{K}-band spectrum of WR 121a also provides a hint about the presence of a companion, in spite of the large extinction towards the source direction \citep{1999AJ....117.1392B}, which suggest that the companion may be a bright object. However, detailed investigation into the matter is necessary. 

An estimate of the intrinsic luminosity from the wind collision in a binary system containing WN7 and O-type dwarf$/$giant$/$supergiant star has been made using the theoretical relations given by \cite{1992ApJ...389..635U}. It has already been mentioned that for all the combinations of WN7 and O-type star, the stellar winds of two binary components interact at sub-terminal speeds either in between the stars or on the surface of the secondary star. Therefore, WR 121a with an orbital period of $\sim$4.1 d can be regarded as one of the very close or close binary systems according to \cite{1992ApJ...389..635U}. However, the predicted intrinsic X-ray luminosity is found to lie in the range of $\sim$0.5$-$6.0 $\times$ 10$^{35}$ erg s$^{-1}$, from the equation (81) of \cite{1992ApJ...389..635U}, where the WR wind is shocking onto the companion surfaces.  On the other hand, when the stellar winds interact in between the stars at sub-terminal speeds, the predicted X-ray luminosity belongs to $\sim$14$-$152  $\times$ 10$^{35}$ erg s$^{-1}$ range, which is one to two orders of magnitude larger than typically seen for CWBs and the value determined here for WR 121a. However, the value of $L_{X}^{int}$ for WR 121a derived from observations is subjected to uncertainties in the distance to the source and the hydrogen column density  used to correct for X-ray absorption. The imprecise estimation of mass-loss rate rates, as well as clumped winds, might also be possible explanations for this discrepancy \citep{1990SvA....34..481C}.

.

\section{Conclusions}\label{conc}
 A deep exploration of the X-ray emission from a WR star WR 121a  using the long-term archival data is presented in this paper. WR 121a is found to be a periodic variable in X-rays with a period of $\sim$4.1 d, where the significant amount of X-rays are originating from the hot plasma heated by the colliding winds.  This periodic variation is also found both in soft and hard energy bands where X-ray flux was increased by $\sim$53\% of minimum flux in hard X-rays. These variations are explained due to the eclipsing of WCR by the companion star in WR 121a system.  The two binary components of WR 121a could not be resolved by \textit{Chandra} and it was detected as a point-like source close to the position of W43 \#1b.  The X-ray spectra of WR 121a below 2.5 keV is heavily absorbed and have been well modelled with two-temperature plasma emission models with the temperature of cool and hot components as $0.98 \pm 0.34 $ and  $3.55 \pm 0.69 $ keV, respectively. The stellar winds from binary components of WR 121a interact at sub-terminal speeds and the wind collision is found to be radiative. The derived values of opening angles for WR 121a suggest that the processes like radiative braking and inhibition must be significantly affecting the wind collision if the companion is an early O-type star. We  encourage more multi-wavelength observations of WR 121a to identify the accurate spectral types of the binary components and to explain the dynamics of winds more precisely in its WCR.

\acknowledgments
 We thank the anonymous referee for the careful reading of the manuscript and giving us his$/$her constructive comments and suggestions. The scientific results reported in this article are based on data obtained from the \textit{Chandra} Data Archive along with observations obtained with \textit{XMM-Newton}, an ESA science mission with instruments and contributions directly funded by ESA Member States and NASA.

\vspace{5mm}
\facilities{\textit{Chandra} and \textit{XMM-Newton}}

\software{CIAO (v4.11; \citealt{2006SPIE.6270E..1VF}),  SAS (v17.0.0; \citealt{2004ASPC..314..759G}), XSPEC (v12.9.1; \citealt{1996ASPC..101...17A}})

\bibliographystyle{aasjournal}
\bibliography{ms}{}

\end{document}